\documentclass[twocolumn]{aastex701}
\usepackage{amsmath}

\newcommand{\code}[1]{\texttt{#1}}

\begin{document}

%\title{A High CO Column on the Inner Side of the HD 163296 CO Snowline}
%\title{The CO Column and $^{13}{\rm C}^{18}{\rm O}$ Optical Depth at the HD 163296 Snowline}
\title{Empirical Constraints on the CO Snowline Transition in HD 163296: The Local CO Column and $^{13}{\rm C}^{18}{\rm O}$ Optical Depth}

\author[0000-0001-8642-1786,sname='Qi']{Chunhua Qi}
\affiliation{Institute for Astrophysical Research, Boston University, 725 Commonwealth Avenue, Boston, MA 02215, USA}
\email[show]{cqi1@bu.edu}

\author[orcid=0000-0003-4902-222X,sname='Ueda']{Takahiro Ueda}
\affiliation{Division of Science, National Astronomical Observatory of Japan, 2-21-1 Osawa, Mitaka-shi, Tokyo 181-8588, Japan}
\email{takahiro.ueda@nao.ac.jp} 

\author[0000-0003-1526-7587,sname='Wilner']{David J. Wilner} 
\affiliation{Center for Astrophysics \textbar\ Harvard \& Smithsonian, Cambridge, MA 02138, USA}
\email{dwilner@cfa.harvard.edu}

\author[0000-0001-9227-5949,sname='Espaillat']{Catherine C. Espaillat} 
\affiliation{Institute for Astrophysical Research, Boston University, 725 Commonwealth Avenue, Boston, MA 02215, USA}
\email{cce@bu.edu}

%% Use the \collaboration command to identify collaborations. This command
%% takes an optional argument that is either a number or the word "all"
%% which tells the compiler how many of the authors above the command to
%% show. For example "\collaboration[all]{(DELVE Collaboration)}" wil include
%% all the authors above this command.
%%
%% Mark off the abstract in the ``abstract'' environment. 
\begin{abstract}
CO isotopologue emission is widely used to infer gas masses and volatile carbon abundances in protoplanetary disks, but converting line emission into a CO column
depends on optical depth, temperature structure, linewidth, and isotope ratios. Inferring CO/H\(_2\) additionally requires an independent constraint on the local hydrogen column.
We use high resolution \(^{13}{\rm C}^{18}{\rm O}\) \(2\!-\!1\) observations of HD~163296 to  
derive a spatially localized empirical constraint on the CO column at the resolved CO snowline edge. We focus on the 70 to 75 au annulus, on the inner, high-column side of the observed profile steepening near 75 au. Using \code{RADEX} slab calculations conditioned on a two dimensional temperature structure, we infer an effective beam-averaged CO column from the absolute integrated intensity. Across representative temperatures, isotope-ratio pairs, and effective local linewidths of 0.30 and 0.50 km s\(^{-1}\), we find \(N_{\rm CO}^{\rm beam}=(1.6\) to \(2.4)\times10^{20}\,{\rm cm^{-2}}\) and \(^{13}{\rm C}^{18}{\rm O}\) line center optical depths \(\tau=0.39\) to 0.97. Thus, even this rare isotopologue is not safely optically thin at the snowline edge.
Adopting \(N_{\rm H2}=2.5\times10^{24}\,{\rm cm^{-2}}\) from a published parametric gas surface density profile gives the conditional abundance \({\rm CO/H_2}=(6.4\) to \(9.5)\times10^{-5}\). A beam-forward radial-profile analysis gives consistent columns, and a physical disk model with a near-canonical warm-layer CO abundance supplies a comparable CO column. The measurement is consistent with the higher C\(^{17}\)O-based MAPS estimate. For the adopted hydrogen column, the inferred CO/H\(_2\) ratio is consistent with a near-canonical abundance on the warm side of the snowline.
\end{abstract}

%% Keywords should appear after the \end{abstract} command. 
%% The AAS Journals now uses Unified Astronomy Thesaurus (UAT) concepts:
%% https://astrothesaurus.org
%% You will be asked to selected these concepts during the submission process
%% but this old "keyword" functionality is maintained in case authors want
%% to include these concepts in their preprints.
%%
%% You can use the \uat command to link your UAT concepts back its source.

%% From the front matter, we move on to the body of the paper.
%% Sections are demarcated by \section and \subsection, respectively.
%% Observe the use of the LaTeX \label
%% command after the \subsection to give a symbolic KEY to the
%% subsection for cross-referencing in a \ref command.
%% You can use LaTeX's \ref and \label commands to keep track of
%% cross-references to sections, equations, tables, and figures.
%% That way, if you change the order of any elements, LaTeX will
%% automatically renumber them.

\section{Introduction}
\label{sec:intro}

The gas mass and volatile inventory of protoplanetary disks set the boundary
conditions for planet formation, but both remain difficult to measure directly.
Molecular hydrogen contains most of the mass but is largely invisible at the
cold temperatures characteristic of planet-forming disks. CO and its
isotopologues are therefore the standard practical tracers of disk gas, because
CO is abundant, chemically stable, and has low-\(J\) rotational transitions
sensitive to cold disk material \citep[e.g.,][]{WilliamsBest2014,
BerginWilliams2017,Zhang2021}. However, converting CO emission into total gas
mass or CO/H\(_2\) abundance is nontrivial: the observed line intensities depend
on optical depth, vertical temperature structure, freezeout, isotope-selective
photodissociation, dust evolution, and chemical conversion of volatile carbon
and oxygen into other reservoirs \citep[e.g.,][]{Visser2009,Miotello2016,
Miotello2017,Bosman2018,Krijt2020,Calahan2021,Zhang2021}.

A growing body of work has suggested that gas phase CO can be depleted relative
to H\(_2\) in Class II disks. In Lupus, Chamaeleon~I, and other nearby
star-forming regions, faint CO isotopologue emission has been interpreted as
evidence for either low gas-to-dust ratios or substantial volatile carbon
depletion \citep{Ansdell2016,Long2017,Miotello2017}. Independent gas mass
constraints from far-infrared HD \(J=1\)--0 observations provide some of the strongest
evidence for this problem: in the few disks with HD detections, CO-based gas masses
can be one to two orders of magnitude smaller than HD-based disk masses, even
after accounting for isotope-selective photodissociation and CO freezeout
\citep{Bergin2013,Favre2013,McClure2016,Calahan2021}. This discrepancy is
commonly interpreted as evidence that carbon- and oxygen-bearing volatiles are
locked into ices, sequestered into large solids, or chemically processed into less volatile reservoirs
\citep[e.g.,][]{Bergin2014,Kama2016,Schwarz2018,Bosman2018,Krijt2020}. 

Complementary constraints have been obtained by jointly modeling CO
isotopologue and N\(_2\)H\(^+\) emission
\citep{Anderson2019,Anderson2022,Trapman2022,Trapman2025}. Gas-phase CO
both competes with N\(_2\) for H\(_3^+\) and efficiently destroys
N\(_2\)H\(^+\), so N\(_2\)H\(^+\) traces regions in which CO has been
removed from the gas. Combining the two tracers therefore helps break the
degeneracy between a low CO abundance and a low total gas mass. Gas masses
derived from joint CO and N\(_2\)H\(^+\) modeling broadly agree with
independent HD- and rotation-curve-based estimates, generally within a
factor of a few, although the results remain sensitive to the thermal
structure, N\(_2\) abundance, and ionization rate.

These studies typically infer CO abundances below the interstellar value
in T~Tauri disks, often by factors of several to tens, but substantially
higher abundances in warmer Herbig disks. In particular,
\citet{Trapman2025} found peak CO abundances near the interstellar value
for their Herbig sample and obtained
\(\log_{10}x_{\rm CO}=-4.28^{+0.17}_{-0.21}\) for HD~163296 by fitting
the spatially integrated 1.3 mm continuum and
\(^{13}\)CO, C\(^{18}\)O, and N\(_2\)H\(^+\) line fluxes. This quantity
is a model parameter describing the peak abundance in the CO-emitting
layer rather than a spatially resolved local abundance.

In the MAPS disks, \citet{Zhang2021} retrieved radial CO gas
distributions using thermochemical modeling, empirical temperature
structures, and C\(^{17}\)O hyperfine fitting. When combined with an assumed
gas to dust ratio of 100, these distributions imply global CO/H\(_2\) abundances
below the interstellar value. The radial behavior is not uniform, however.
The three T~Tauri disks show substantial CO underabundances over broad radial
ranges, whereas HD~163296 and MWC~480 have lower inferred CO abundances 
in their outer disks followed by a rapid inward rise across their midplane
snowlines. For HD~163296, the C\(^{17}\)O-based result approaches the
interstellar CO/H\(_2\) value inside the snowline
\citep{Zhang2021}.

Taken together, these results demonstrate that global or disk-averaged CO
depletion estimates should not be extrapolated directly to the warm side of 
a resolved snowline. More generally, faint CO isotopologue emission does not
uniquely imply disk wide elemental carbon depletion. The observable CO
isotopologue column can also be reduced by the vertical density and
temperature structure, isotope-selective photodissociation, freezeout, and
grain-surface processing
\citep{Ruaud2022,Deng2023,Deng2025,Deng2026}. In self-consistent DiskMINT
models, for example, grain-surface conversion of CO ice into more tightly
bound CO$_2$-rich ice can reduce the gas-phase CO isotopologue columns
without imposing an ad hoc, disk-wide elemental depletion factor
\citep{Ruaud2022,Deng2026}.

[C~I] observations provide an important complementary diagnostic of the volatile
carbon reservoir. In
particular, the ACA [C~I] \(1-0\) survey of large Lupus T~Tauri disks by
\citet{Pascucci2023} found that most \(\sim1\)--3 Myr disks with gas radii
\(\gtrsim200\) au are consistent with thermochemical models that adopt
interstellar carbon and oxygen abundances \citep{Ruaud2022}. 
Analyses of warmer Herbig disks likewise suggest that the discrepancy
between HD-based and CO-based gas masses may be reduced or absent in some
systems, and that CO isotopologue emission can remain optically thick enough
for optically thin mass estimates to underestimate the gas column
\citep{Kama2020,Stapper2024}. These results emphasize that inferences of CO depletion
depend on disk population, size, thermal structure, optical depth, and tracer selection. 
As demonstrated for HD~163296 by \citet{QiWilner2024}, a sharp
radial decline in rare CO isotopologue emission need not imply a
globally depleted CO/H\(_2\) abundance but may instead trace the physical
CO-column transition associated with the snowline.

The inferred CO abundance therefore depends on disk radius, tracer
selection, and the adopted physical and chemical structure. This motivates
a clear separation between quantities constrained more directly by the
observations and the physical interpretations assigned to them.

HD~163296 provides a critical test case. It is a massive Herbig disk with
well-resolved continuum substructures, extensive CO isotopologue observations,
and a sharply defined CO snowline transition. 
An important earlier constraint
was provided by \citet{Booth2019}, who reported the first detection of
\(^{13}{\rm C}^{17}{\rm O}\) \(3\!-\!2\) in a protoplanetary disk.
Their radiative transfer model underpredicted the disk integrated
\(^{13}{\rm C}^{17}{\rm O}\) intensity and required a factor of 3.5
increase in the global CO gas mass. They also found that C\(^{18}\)O is
optically thick inside the snowline, whereas
\(^{13}{\rm C}^{17}{\rm O}\) remains optically thin. This result provided
early model-based evidence for a substantial CO reservoir in HD~163296, although the
relative coarse spatial resolution (\(0\farcs87\times0\farcs51\), corresponding to a characteristic linear
resolution of \(\sim70\) au) did not isolate the CO column at the resolved snowline edge.

The MAPS analysis subsequently showed that HD~163296 CO column is radially structured. In
particular, \citet{Zhang2021} found that C\(^{17}\)O \(1-0\)-based analysis gives CO columns
approximately \(2\)--\(6\) times higher than C\(^{18}\)O \(2-1\)-based result inside \(\sim100\) au. 
The higher C\(^{17}\)O-based column
also better reproduced the available disk-integrated
\(^{13}{\rm C}^{18}{\rm O}\) spectrum, motivating higher-resolution
observations of this tracer inside \(100\) au.
The recent spatially resolved \(^{13}{\rm C}^{18}{\rm O}\) \(2-1\)
observations of HD~163296 \citep{Armitage2026} provide exactly this
opportunity. \citet{Armitage2026} used these observations to infer centrally
peaked CO enhancement and to discuss pebble-drift histories in HD~163296 and
MWC~480. Here, the same HD~163296 data are used for a complementary purpose: 
 to constrain the local beam-averaged CO column and optical depth on
the high-column side of the resolved snowline transition, and to evaluate the
corresponding conditional CO/H\(_2\) abundance.

The focus on the high-column side of the snowline transition 
is motivated by the observational snowline diagnostic
identified by \citet{QiWilner2024,Qi2026}. Using the same \(^{13}{\rm C}^{18}{\rm O}\)
\(2-1\) data, \citet{Qi2026} showed that the radial derivative of the
integrated-intensity profile has a clear local minimum near \(0\farcs75\),
coincident with the sharp CO snowline transition traced by other rare CO
isotopologues \citep{QiWilner2024}. They further showed that same-transition
rare-isotopologue ratios rise rapidly across the snowline and flatten in the
outer disk, thereby empirically identifying an optically thin region suitable for
isotopic-ratio measurements. Their analysis established the \(0\farcs75\)
derivative feature as an observational marker of a sharp CO-column transition,
but did not determine the absolute local \(N_{\rm CO}\), optical depth, or
CO/H\(_2\) abundance on its high-column side. The present analysis constrains the local CO column and
\(^{13}{\rm C}^{18}{\rm O}\) optical depth at the resolved HD~163296
snowline edge; the corresponding CO/H\(_2\) ratio is evaluated separately
and remains conditional on the adopted hydrogen column.

In this paper, we use the high-resolution \(^{13}{\rm C}^{18}{\rm O}\)
\(2-1\) profile to constrain the CO column and conditional CO/H\(_2\)
abundances on the inner, high-column side of the HD~163296
CO snowline transition. 
We first infer an effective local beam-averaged \(N_{\rm CO}\) and
\(^{13}{\rm C}^{18}{\rm O}\) optical depth from the absolute integrated
intensity, using a \code{RADEX} slab calculation conditioned on a two-dimensional
temperature structure. We then use a beam-forward radial-profile MCMC analysis to
test the effects of the radial beam coupling on the local inference. Finally,
we use a physical disk model that was not
fitted to the \(^{13}{\rm C}^{18}{\rm O}\) radial profile to perform a column-budget consistency check, determining whether plausible gas surface densities and  a self-consistent temperature structure, together with a near-canonical warm CO abundance, can supply the high CO column inferred on the inner side of the snowline transition.

Section~\ref{sec:data} describes the observations and radial-profile
extraction. Section~\ref{sec:modeling} presents the local beam-averaged
inversion, the beam-forward consistency check, and the physical disk model used for the column-budget comparison.
Section~\ref{sec:results} presents the inferred CO column, optical depth, radial
decline, physical column-budget check, and conditional CO/H\(_2\) abundance. We
show that the absolute \(^{13}{\rm C}^{18}{\rm O}\) intensity requires a high local
CO column at 70--75 au, the inner side of the HD~163296 CO snowline transition. 
For the adopted local gas column, the inferred CO column corresponds to a
near-canonical CO/H\(_2\) abundance.

\section{Observations and Radial Profile}
\label{sec:data}

We use the high-resolution \(^{13}{\rm C}^{18}{\rm O}\) \(J=2-1\)
observations of HD~163296 previously analyzed by \citet{Qi2026}. The
observations were obtained in ALMA project 2021.1.00899.S and are described in
detail by \citet{Armitage2026}. The image cube used here has a synthesized beam
of \(0.274''\times0.246''\), is corrected for the JvM effect\citep{Czekala2021},
and is primary-beam corrected. The continuum was subtracted in the visibility
domain before line imaging. No explicit correction for dust absorption or scattering 
is applied to the continuum-subtracted line intensity. The emission is consistent with Keplerian
rotation, allowing a deprojected radial profile to be extracted from the cube.

We extract a Keplerian-stacked radial integrated-intensity profile with \texttt{GoFish} \citep{Teague2019},
using \(i=46.7^\circ\), \({\rm PA}=313.3^\circ\), and
\(v_{\rm sys}=5.76~{\rm km~s^{-1}}\). The fiducial profile is integrated over
a velocity-offset window of \(\pm8~{\rm km~s^{-1}}\) relative to the projected
Keplerian velocity. This window is sufficiently broad to recover the line flux
while avoiding unnecessary noise from line-free channels. Tests with
\(\pm4\), \(\pm6\), \(\pm8\), and \(\pm10~{\rm km~s^{-1}}\) windows show that
the \(\pm8~{\rm km~s^{-1}}\) profile is effectively flux-converged relative to
the \(\pm10~{\rm km~s^{-1}}\) extraction. We therefore adopt the
\(\pm8~{\rm km~s^{-1}}\) profile as the fiducial radial profile used below.
Unless otherwise stated, we use the JvM-corrected, primary-beam-corrected
cube for the quantitative analysis. To test the sensitivity to residual
scaling, we repeated the full radial-profile extraction using the standard
restored cube without the JvM correction. The two products use the same
visibilities, weighting, channelization, CLEAN model, restoring beam, mask,
velocity interval, disk geometry, and radial bins; only the scaling of the
residual image differs.

\begin{figure*}[ht!]
\centering
\includegraphics[width=\linewidth]{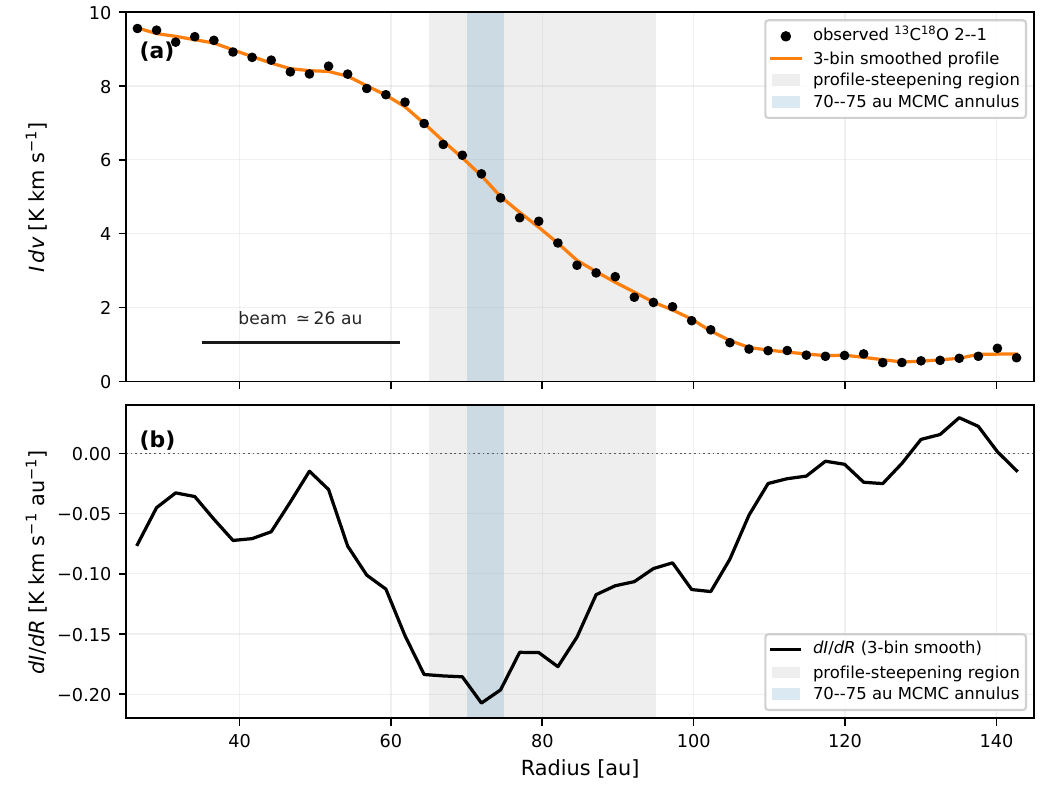}
\caption{
Observed \(^{13}{\rm C}^{18}{\rm O}\) \(J=2-1\) radial profile of HD~163296
used to define the column-density inference region.
(a) Keplerian-stacked radial integrated-intensity profile extracted with the
fiducial \(\pm8~{\rm km~s^{-1}}\) velocity-offset window. Black points show the
observed radial profile, and the orange curve shows the 3-bin smoothed profile
used only for the derivative visualization. The horizontal bar marks the
representative radial beam FWHM,
\(\sqrt{0.274''\times0.246''}\simeq26\) au.
(b) Radial derivative of the smoothed profile. The gray band marks the broad
profile-steepening region, \(R\simeq65\)--95 au, rather than a single fitted
snowline radius. The blue band marks the 70--75 au annulus used for the MCMC
inference of \(N_{\rm CO}\), \(\tau_{^{13}{\rm C}^{18}{\rm O}}\), and
CO/H\(_2\). This annulus lies within the inner, high-intensity side of the
observed steepening region and is chosen to provide a conservative estimate of
the CO abundance immediately interior to the CO snowline transition.
}
\label{fig:profile_derivative}
\end{figure*}

Figure~\ref{fig:profile_derivative} shows the
\(^{13}{\rm C}^{18}{\rm O}\) radial profile and its radial derivative. The
profile declines smoothly at small radii and then steepens across the CO
snowline region. We use the derivative only to identify the broad radial zone
where the observed, beam-convolved profile steepens; we do not assign a unique
snowline radius to the most negative derivative point. The precise location of
that point can depend on the velocity-integration window, radial binning,
derivative smoothing, and beam convolution. We therefore mark
\(R\simeq65\)--95 au as a conservative profile-steepening region, comparable
to the beam-convolved width of the feature, rather than as a resolved
transition width.

The 70--75 au annulus used for the fiducial column-density inference lies on
the inner, high-intensity side of this profile-steepening region. We choose
this annulus because it provides a useful compromise for an empirical
\(N_{\rm CO}\) measurement. Moving outward samples gas increasingly affected
by CO freezeout, while moving substantially inward leads to brighter
\(^{13}{\rm C}^{18}{\rm O}\) emission that is expected to become more
optically thick, making the emergent intensity less sensitive to the total CO
column. The 70--75 au annulus is therefore close enough to the derivative
feature to probe the snowline edge, but still lies on the high-column side
where the abundance can be compared most directly with the canonical
CO/H\(_2\) value using an explicit optical-depth correction. If this annulus is
already partially affected by freezeout, the inferred CO/H\(_2\) abundance is
conservative relative to the fully warm-side abundance immediately interior to
the transition.

\section{CO Column and Optical-Depth Inference}
\label{sec:modeling}

The radial profile and derivative in Figure~\ref{fig:profile_derivative}
identify where the observed, beam-convolved \(^{13}{\rm C}^{18}{\rm O}\)
\(2\!-\!1\) emission steepens, but they do not by themselves determine
\(N_{\rm CO}\). The measured profile is convolved with the
\(0.274''\times0.246''\) synthesized beam, so a direct inversion of an
annular intensity gives an effective beam-averaged column rather than a
deconvolved intrinsic \(N_{\rm CO}(R)\). We therefore separate the analysis
into three complementary components. First, we perform a local beam-averaged inversion of the
absolute integrated intensity in selected annuli, using the 70--75 au annulus
as the fiducial abundance constraint. Second, we use a beam-forward
radial-profile MCMC to test whether emission from neighboring radii, coupled
by the synthesized beam, changes this local conclusion. In this
calculation, trial intrinsic \(N_{\rm CO}(R)\) profiles are converted to
\(^{13}{\rm C}^{18}{\rm O}\) emission, convolved with the ALMA beam, and
compared with the observed radial profile. We use it as a
consistency check, not as a unique deconvolved measurement of
\(N_{\rm CO}(R)\). Third, we use an independent physical disk model for 
a column-budget check, checking whether plausible gas and dust surface densities, 
together with a self-consistent temperature structure and near-canonical warm CO 
abundance, can supply the high CO column inferred from the local inversion.

\subsection{Local Beam-Averaged Inversion}
\label{subsec:beamavg_inversion}

For an annulus \(R_0\)--\(R_1\), we treat the observed radial profile
intensity as a beam-convolved quantity, \(I_{\rm obs}^{\rm beam}\), and solve
for the uniform-slab CO column that reproduces this absolute integrated
intensity. The inferred quantity,
\(N_{\rm CO}^{\rm beam}(R_0\text{--}R_1)\), is therefore an effective
beam-averaged column; it is not a deconvolved intrinsic measurement of
\(N_{\rm CO}(R)\).

For each trial \(N_{\rm CO}^{\rm beam}\), we convert the total CO column to a
\(^{13}{\rm C}^{18}{\rm O}\) column using
\(N(^{13}{\rm C}^{18}{\rm O}) =
N_{\rm CO}^{\rm beam}/[(^{12}{\rm C}/^{13}{\rm C})
(^{16}{\rm O}/^{18}{\rm O})]\). We consider two isotope-ratio choices: local
ISM values, \(^{12}{\rm C}/^{13}{\rm C}=69\) and
\(^{16}{\rm O}/^{18}{\rm O}=557\), and the HD~163296 median values from
\citet{Qi2026}, \(^{12}{\rm C}/^{13}{\rm C}=75.3\) and
\(^{16}{\rm O}/^{18}{\rm O}=599\). These choices bracket the isotopic
conversion uncertainty relevant for deriving \(N_{\rm CO}\) from
\(^{13}{\rm C}^{18}{\rm O}\).

We compute the \(^{13}{\rm C}^{18}{\rm O}\) \(2\!-\!1\) integrated intensity
with \code{RADEX} through the \code{SpectralRadex} interface \citep{vanderTak2007,
Holdship2021}. 
The gas kinetic temperature is taken
from the two-dimensional thermal structure inferred for HD~163296 by
\citet{FehrAndrews2025}. We denote the cylindrical radius and height above
the disk midplane by \(R\) and \(z\), respectively, and define the
dimensionless height
\(\zeta\equiv |z|/R\). In each radial annulus, the vertical temperature
profile is parameterized as
\begin{equation}
T(R,z)=
\begin{cases}
T_{\rm mid}(R),
& \zeta \leq a_z(R), \\[3pt]
T_{\rm mid}(R)
\exp\!\left\{
\gamma(R)\left[\zeta-a_z(R)\right]
\right\},
& \zeta > a_z(R).
\end{cases}
\label{eq:fehr_temperature}
\end{equation}
where \(T_{\rm mid}(R)\) is the midplane temperature,
\(a_z(R)R\) marks the height at which the vertically isothermal midplane
transitions to an exponential temperature rise, and \(\gamma(R)\) controls
the vertical temperature gradient. The parameters
\(T_{\rm mid}\), \(a_z\), and \(\gamma\) were inferred independently in
radial annuli from the emitting surfaces and brightness temperatures of 
multiple optically thick CO isotopologue emission \citep{FehrAndrews2025}.

For the fiducial inversion, we use the posterior-median midplane
temperature, \(T(R,0)=T_{\rm mid}(R)\), because the selected 70--75 au
annulus lies near the midplane CO snowline and the 
\(^{13}{\rm C}^{18}{\rm O}\) emission is
expected to receive substantial contributions from
the dense, low-altitude CO-bearing column. We also
evaluate the temperature structure at \(\zeta=0.1\) and 0.2 as sensitivity
tests that bracket uncertainty in the effective excitation temperature.
These heights are not assumed or measured
\(^{13}{\rm C}^{18}{\rm O}\) emitting surfaces. For an optically thin or
marginally optically thick transition, the intensity integrates emission
through the CO-bearing vertical column rather than emerging from a unique
\(\tau=1\) surface. Empirical temperatures derived from bright CO emitting
surfaces, including the MAPS analysis of \citet{Calahan2021}, provide
important constraints on the upper molecular layers but are less directly
matched to the low-altitude region relevant to the present local column
Inversion.

We adopt \(n_{\rm H2}=10^7~{\rm cm^{-3}}\), include ortho- and para-H\(_2\)
collision partners with an ortho-to-para ratio of 3:1, set
\(T_{\rm bg}=2.73~{\rm K}\), and use an escape-probability slab geometry.
Tests with \(\log n_{\rm H2}=6\)--8 change the inferred \(N_{\rm CO}\) at
70--75 au by less than one percent, indicating that the
\(^{13}{\rm C}^{18}{\rm O}\) \(2\!-\!1\) excitation is effectively
thermalized for the relevant conditions.

The linewidth used in \code{RADEX} is the effective local FWHM that
sets the line-center optical depth and escape probability, rather than the
macroscopically broadened width of an annular stacked spectrum. For
\(^{13}{\rm C}^{18}{\rm O}\), thermal broadening gives
\(\Delta v_{\rm th}\simeq0.18\)--0.24 \({\rm km~s^{-1}}\) for
\(T=20\)--40 K. We adopt
\(\Delta v=0.30~{\rm km~s^{-1}}\) as the fiducial effective linewidth. This
value is only modestly broader than the thermal expectation and allows for
unresolved local nonthermal broadening and sub-beam velocity gradients. The 
excess over the thermal width is not interpreted as a measurement of turbulence.

Velocity gradients along the line of sight can redistribute the opacity over
velocity and thereby photon escape, while unresolved Keplerian gradients
across the synthesized beam broaden the observed annular spectrum. These
macroscopic contributions are not represented explicitly in the uniform
\code{RADEX} slab, and the stacked-spectrum linewidth should therefore 
not be used directly as its local linewidth. We 
also calculate models with
\(\Delta v=0.50~{\rm km~s^{-1}}\) to provide a conservative systematic
bracket on this effective local parameter.

% Diagnostic test

For each fixed combination of nuisance parameters,
\(\{\zeta,\Delta v,R_{12/13},R_{16/18}\}\), \code{RADEX} provides the
modeled integrated intensity
\(I_{\rm model}(N_{\rm CO}^{\rm beam})\) and the corresponding line-center
optical depth \(\tau_{^{13}{\rm C}^{18}{\rm O}(2-1)}^{\rm beam}\). 
We consider the two isotope-ratio pairs introduced above:
the local ISM values
\((R_{12/13},R_{16/18})=(69,557)\) and the QiMed values
\((75.3,599)\). 
The local likelihood is
\[
\chi^2_{\rm beam} =
\left[
\frac{I_{\rm obs}^{\rm beam}
-I_{\rm model}(N_{\rm CO}^{\rm beam})}{\sigma_I}
\right]^2 ,
\]
where \(\sigma_I\) includes the radial-profile uncertainty and a 10\% absolute
calibration term added in quadrature. We adopt a uniform prior in
\(\log N_{\rm CO}^{\rm beam}\) over a broad range and compute the
one-parameter posterior on a dense grid. For this single-parameter inversion,
the grid posterior is equivalent to an MCMC posterior but avoids sampler noise.

The fiducial abundance estimate uses the 70--75 au annulus, which lies on the
inner, high-intensity side of the profile-steepening region identified in
Figure~\ref{fig:profile_derivative}. We also apply the same inversion to
non-overlapping 10 au annuli to test whether the effective CO column declines
outward across the transition. Because the representative radial beam FWHM is
\(\simeq26\) au, these annuli are not independent measurements of the
intrinsic 10 au-scale profile; they provide a transparent beam-averaged check
on the radial trend.

\subsection{Beam-Forward Radial-Profile Consistency Check}
\label{subsec:beamforward_consistency}

The local beam-averaged inversion intentionally does not deconvolve the radial
structure. To test whether neighboring emission coupled by the synthesized beam
affects the 70--75 au result, we also perform a beam-forward radial-profile
MCMC. The intrinsic CO column is represented by free nodes
\(\{R_i,\log N_{{\rm CO},i}\}\), which are interpolated with a
shape-preserving piecewise cubic Hermite interpolation,
\(\log N_{\rm CO}(R)={\rm PCHIP}[\log N_{{\rm CO},i},R_i]\). PCHIP provides a
smooth profile while avoiding the spurious overshoots that can occur with
standard cubic splines. The nodes are spaced more finely across the observed
profile-steepening region and more coarsely elsewhere.

This flexible profile is used only as a nuisance representation of the radial
structure needed for beam convolution. It does not impose an analytic snowline
function, a prescribed depletion profile, or a global thermochemical abundance
structure. We therefore do not interpret the detailed sampled profile as a
unique, model-independent \(N_{\rm CO}(R)\) measurement. The local quantity
reported from this calculation is the marginalized beam-forward column
\(N_{\rm CO}^{\rm bf}(70\text{--}75~{\rm au})\), together with the
corresponding \(\tau_{^{13}{\rm C}^{18}{\rm O}(2-1)}\).

For each trial \(N_{\rm CO}(R)\), we use the same \code{RADEX} conversion
described above to compute the intrinsic
\(^{13}{\rm C}^{18}{\rm O}\) \(2\!-\!1\) intensity and optical-depth
profiles. The intrinsic intensity profile is projected into the disk image
plane assuming axisymmetry and the same geometry used for the \code{GoFish}
extraction, convolved with the observed ALMA beam, and re-extracted using the
observed radial bins. We evaluate the agreement between the model and
the data using the radial-profile chi-square statistic,
\[
\chi^2_{\rm prof} =
\sum_j
\left[
\frac{I_{\rm obs}(R_j)-I_{\rm model}(R_j)}
{\sigma_I(R_j)}
\right]^2 .
\]
Because neighboring radial bins are correlated by the beam, this is an
effective profile likelihood rather than a set of independent radial
measurements.

To suppress unresolved node-to-node ringing without imposing a transition
shape, we add a weak curvature penalty,
\[
\chi^2_{\rm reg} =
\sum_i
\left[
\frac{\Delta^2\log N_{{\rm CO},i}}{\sigma_{\rm smooth}}
\right]^2 ,
\]
where \(\Delta^2\log N_{{\rm CO},i}\) is the second difference of the node
sequence. This smoothness prior does not prescribe a transition radius, width,
contrast, or monotonic radial slope. The sampled posterior is proportional to
\(\ln p=-\frac{1}{2}(\chi^2_{\rm prof}+\chi^2_{\rm reg})+\ln p_{\rm bounds}\).

We adopt independent uniform bounds on the column at every node,
\(
17 \leq
\log_{10}\!\left(\frac{N_{{\rm CO},i}}{{\rm cm}^{-2}}\right)
\leq 22.
\)
Thus, \(\ln p_{\rm bounds}=0\) when all node columns lie within these
limits and \(-\infty\) otherwise.

The beam-forward MCMC uses the same eight fixed nuisance-parameter cases as
the local inversion: \(\zeta=0.10\) and 0.20,
\(\Delta v=0.30\) and \(0.50~{\rm km~s^{-1}}\), and the two isotope-ratio
choices above. Each production run uses 64 walkers and 15000 steps, with the
first 2500 steps discarded as burn-in and the retained chain thinned by 30. We
verify convergence using the integrated autocorrelation time and acceptance
fraction.

The CO/H\(_2\) abundance is computed after the column inference by dividing
the inferred CO column by the adopted local \(N_{\rm H2}\); the H\(_2\) column
is not sampled in either likelihood. The primary results are the
beam-averaged \(N_{\rm CO}^{\rm beam}\), optical depth, and conditional
CO/H\(_2\) abundance. The beam-forward \(N_{\rm CO}^{\rm bf}\) values are used
only as a consistency check on radial beam-coupling and as a descriptive
diagnostic of the broader CO-column transition.

\subsection{Physical Disk Model for the Column-Budget Check} \label{subsec:physical_model}

To assess whether the empirically inferred CO column is physically plausible, we construct an independent disk-structure model for HD~163296. This model is not fit to the\(^{13}{\rm C}^{18}{\rm O}\) data and is not used in the empirical \(N_{\rm CO}\) inference. Instead, it provides a column-budget check on whether plausible gas and dust surface densities, together with a self-consistent temperature structure, contain enough warm, shielded CO to match the column inferred from the local line inversion.

We first estimate the radial dust surface density and maximum grain size based on the multi-wavelength dust continuum images at $\lambda=0.88$, 1.3, 2.1, 2.9, 3.3 and 8.6 mm from \citet{Guidi2022}.
We perform an MCMC fit to the azimuthally-averaged radial intensity profiles using an analytical expression for the emergent intensity that includes scattering (e.g., \citealt{Sierra+2019}).
In this fit, we assume a dust size distribution with a power-law index of $-3.5$, with a porosity of 0.9 and optical constants based on the DSHARP mixture \citep{Birnstiel2018}, replacing refractory organics with amorphous carbon \citep{Zubko+96}.
This replacement is motivated by recent multi-wavelength studies showing
that amorphous-carbon-rich mixtures can better reproduce resolved
millimeter-to-centimeter continuum spectra than the default DSHARP
composition \citep{Zagaria2025}.
Population-synthesis studies provide complementary evidence that opacity prescriptions with higher millimeter absorption efficiencies reproduce observed disk continuum properties better than the default DSHARP opacity \citep{Delussu2024}.

These assumptions define the dust opacity model used for the continuum-based \(\Sigma_{\rm dust}(R)\) and \(a_{\max}(R)\) estimates.
Using the inferred radial dust distribution, we construct a two-dimensional dust density structure and compute the disk temperature structure with RADMC-3D \citep{RADMC}. To convert the radial profiles into a two-dimensional distribution, we divide the dust population into four size bins per decade in grain size and account for size-dependent dust settling in each bin, assuming a turbulence strength of $\alpha = 3\times10^{-3}$. In addition to the dust population inferred from the continuum emission, we include a small-grain population that is fully coupled to the gas and has a total mass equal to 3\% of the continuum-derived dust mass. Because the dust scale height depends on the thermal structure, we iteratively recompute the dust distribution and temperature structure until a self-consistent solution is obtained. 

For the gas surface density, we adopt the best-fit HD~163296 disk model from the MAPS analysis \citep{Zhang2021}. We consider both this smooth gas profile and a gapped variant, constructed by modifying the gas surface density by three Gaussian gap functions, $(1-\delta_i\exp[-((r-r_i)/w_i)^2])$, with $(r_i,\delta_i,w_i)=(55~{\rm au},0.6,5~{\rm au})$, $(85~{\rm au},0.5,10~{\rm au})$, and $(145~{\rm au},0.5,20~{\rm au})$. 
The gap centers were chosen based on the locations of the
prominent dust continuum gaps; these locations are also close to
previously reported kinematic substructures in HD~163296
\citep{Teague+2018,Izquierdo2022,Alarcon2022}.
The widths and depths were chosen phenomenologically and were not
fitted to the CO observations; this variant is intended only as a
sensitivity test and not as a reproduction of the CO-gap model of
\citet{Zhang2021}.
The corresponding two-dimensional gas density and temperature structures are used only for the physical column-budget comparison in  Section~\ref{subsec:physical_budget}.

\section{Results}
\label{sec:results}

We now combine the local intensity inversion, beam-forward modeling, and
physical column budget check to assess the CO column on the high-column side of
the HD~163296 snowline transition. Section~\ref{subsec:nco_70_75} shows
that the absolute \(^{13}{\rm C}^{18}{\rm O}\) brightness at 70--75 au
requires a high beam-averaged \(N_{\rm CO}\) and a line that is not safely
optically thin. Section~\ref{subsec:beamforward_results} tests whether this conclusion
is altered by radial beam coupling. Section~\ref{subsec:physical_budget}
then evaluates whether independent physical disk models contain enough warm,
shielded CO to supply the inferred column. The remaining
subsections discuss linewidth and \(N_{\rm H2}\) systematics, compare with MAPS, and
summarize the requirements for a more fully empirical \(N_{\rm CO}(R)\) reconstruction.

\subsection{Local Beam-Averaged CO Column and Radial Decline}
\label{subsec:nco_70_75}

Figure~\ref{fig:profile_derivative} identifies the radial zone where the
beam-convolved \(^{13}{\rm C}^{18}{\rm O}\) \(2\!-\!1\) emission steepens.
We first convert the absolute intensity in this region into an effective
beam-averaged CO column, \(N_{\rm CO}^{\rm beam}\), defined as the
uniform-slab CO column required to reproduce the observed
\(^{13}{\rm C}^{18}{\rm O}\) brightness for the adopted temperature structure,
local linewidth, isotope ratios, and density. This calculation does not fit or
deconvolve a full radial \(N_{\rm CO}(R)\) profile.

The fiducial abundance estimate uses the 70--75 au annulus, on the inner,
high-intensity side of the profile-steepening region. The inferred
\(N_{\rm CO}^{\rm beam}(70\text{--}75~{\rm au})\) is appropriate for the
observed \(0\farcs274\times0\farcs246\) beam and should not be interpreted as
a deconvolved intrinsic 5 au-wide measurement. The representative radial beam
FWHM is \(\simeq26\) au, so neighboring annuli are beam-correlated.

\begin{figure*}[ht!]
\centering
\includegraphics[width=\linewidth]{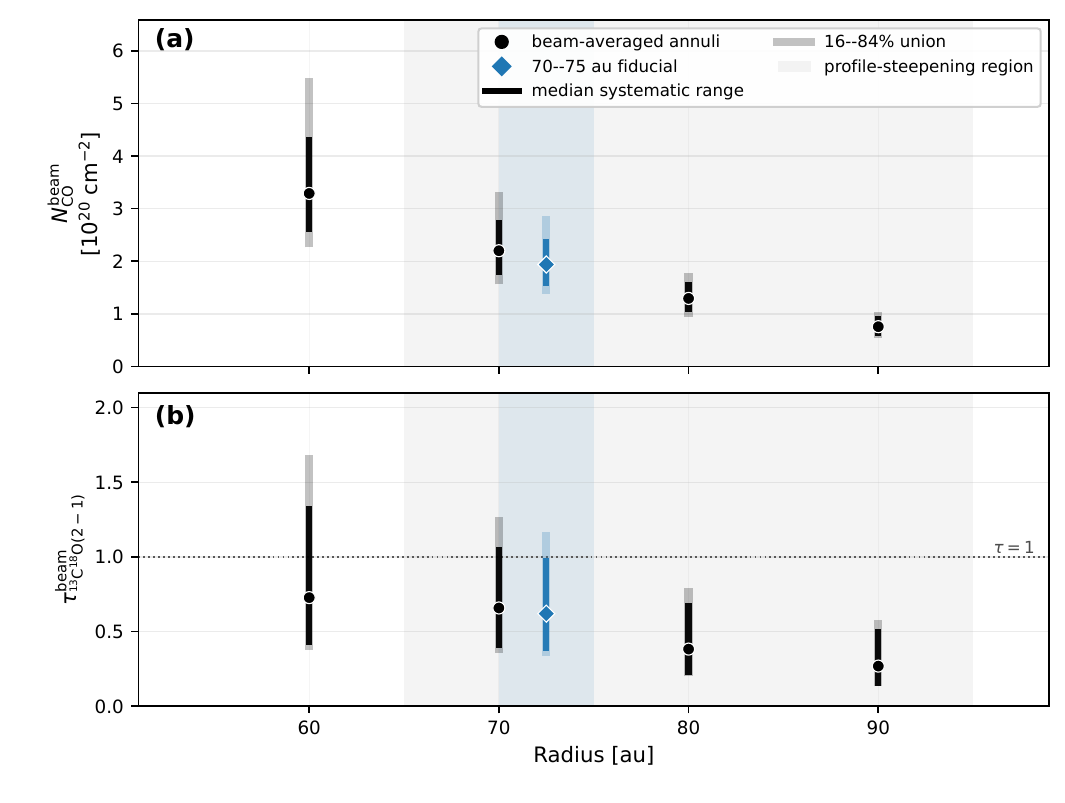}
\caption{
Local beam-averaged CO-column inversion from the observed
\(^{13}{\rm C}^{18}{\rm O}\) \(2\!-\!1\) integrated intensity.
(a) Effective beam-averaged CO column, \(N_{\rm CO}^{\rm beam}\), in
non-overlapping annuli across the profile-steepening region. Black circles
show the median across the fixed nuisance-parameter cases; thick vertical bars
show the range of posterior medians across cases, and light vertical bars show
the union of the 16--84 percentile posterior intervals. The blue diamond marks
the fiducial 70--75 au annulus.
(b) Corresponding beam-averaged \(^{13}{\rm C}^{18}{\rm O}\) \(2\!-\!1\)
optical depth. The dotted line marks \(\tau=1\). The gray band marks the
profile-steepening region from Figure~\ref{fig:profile_derivative}, and the
blue band marks the fiducial annulus. These are beam-averaged quantities, not
deconvolved intrinsic \(N_{\rm CO}(R)\) measurements.
}
\label{fig:beamavg_nco_tau}
\end{figure*}

Figure~\ref{fig:beamavg_nco_tau} and Table~\ref{tab:beamavg_nco_compact}
summarize the local inversion. Across the eight fixed nuisance-parameter
cases, spanning two local linewidths, two isotope-ratio choices, and two
representative emitting heights, the fiducial annulus gives
\[
N_{\rm CO}^{\rm beam}(70\text{--}75~{\rm au})
=
(1.6\text{--}2.4)\times10^{20}~{\rm cm^{-2}},
\]
where the range denotes the posterior medians across the fixed cases. The
corresponding line-center optical depths are
\[
\tau_{^{13}{\rm C}^{18}{\rm O}(2-1)}^{\rm beam}=0.39\text{--}0.97 .
\]
Thus \(^{13}{\rm C}^{18}{\rm O}\) \(2\!-\!1\) is optically thin to marginally
thick on the high-column side of the transition, and a purely optically thin
conversion is not adequate for all cases.

Applying the same inversion to neighboring non-overlapping annuli tests whether
the profile steepening is accompanied by an outward decrease in the effective
CO column. Across the fixed nuisance-parameter cases,
\(N_{\rm CO}^{\rm beam}\) decreases from
\((2.6\text{--}4.3)\times10^{20}~{\rm cm^{-2}}\) at 55--65 au to
\((1.8\text{--}2.7)\times10^{20}~{\rm cm^{-2}}\) at 65--75 au,
\((1.1\text{--}1.5)\times10^{20}~{\rm cm^{-2}}\) at 75--85 au, and
\((0.64\text{--}0.90)\times10^{20}~{\rm cm^{-2}}\) at 85--95 au. The
effective column therefore declines monotonically across the
profile-steepening region; even from 65--75 to 75--85 au, the decrease is a
factor of \(\sim1.5\)--1.8 despite beam smearing. These annuli are not
independent measurements of the intrinsic 10 au-scale structure and should not
be used to infer a resolved transition width. They do, however, show that the
derivative feature in Figure~\ref{fig:profile_derivative} is accompanied by an
outward decrease in the beam-averaged CO column, placing the fiducial
70--75 au annulus on the high-column side of the transition.

\begin{deluxetable}{lccccc}
\tabletypesize{\scriptsize}
\tablecaption{Beam-averaged CO-column constraints at 70--75 au\label{tab:beamavg_nco_compact}}
\tablehead{
\colhead{Isotopes} &
\colhead{\(\zeta\)} &
\colhead{\(\Delta v\)} &
\colhead{\(N_{\rm CO}^{\rm beam}\)} &
\colhead{\(\tau_{13{\rm C}^{18}{\rm O}}^{\rm beam}\)} &
\colhead{CO/H\(_2\)} \\
\colhead{} &
\colhead{} &
\colhead{km s\(^{-1}\)} &
\colhead{\(10^{20}\,{\rm cm^{-2}}\)} &
\colhead{} &
\colhead{\(10^{-5}\)}
}
\startdata
ISM   & 0.10 & 0.30 & \(2.02^{+0.41}_{-0.34}\) & \(0.97\) & \(8.1^{+1.7}_{-1.4}\) \\
QiMed & 0.10 & 0.30 & \(2.37^{+0.49}_{-0.40}\) & \(0.97\) & \(9.5^{+1.9}_{-1.6}\) \\
ISM   & 0.10 & 0.50 & \(1.60^{+0.23}_{-0.21}\) & \(0.46\) & \(6.4^{+0.9}_{-0.8}\) \\
QiMed & 0.10 & 0.50 & \(1.87^{+0.27}_{-0.25}\) & \(0.46\) & \(7.5^{+1.1}_{-1.0}\) \\
ISM   & 0.20 & 0.30 & \(1.96^{+0.35}_{-0.31}\) & \(0.78\) & \(7.8^{+1.4}_{-1.2}\) \\
QiMed & 0.20 & 0.30 & \(2.30^{+0.41}_{-0.36}\) & \(0.78\) & \(9.2^{+1.6}_{-1.4}\) \\
ISM   & 0.20 & 0.50 & \(1.63^{+0.22}_{-0.21}\) & \(0.39\) & \(6.5^{+0.9}_{-0.8}\) \\
QiMed & 0.20 & 0.50 & \(1.92^{+0.26}_{-0.25}\) & \(0.39\) & \(7.7^{+1.0}_{-1.0}\) \\
\enddata
\tablecomments{
Intervals are the 16--84 percentile posterior ranges for each fixed
nuisance-parameter case, including the radial-profile uncertainty and a
10\% absolute calibration term. ``ISM'' adopts
\(^{12}{\rm C}/^{13}{\rm C}=69.0\) and
\(^{16}{\rm O}/^{18}{\rm O}=557\); ``QiMed'' adopts 75.3 and 599.
CO/H\(_2\) assumes \(N_{\rm H2}=2.5\times10^{24}\,{\rm cm^{-2}}\) and scales
inversely with the adopted H\(_2\) column.
}
\end{deluxetable}

Repeating the 70--75 au inversion with only the statistical profile
uncertainty, omitting the 10\% absolute calibration term, narrows the posterior
intervals but changes the median columns by only a few percent. The high
column is therefore set by the absolute \(^{13}{\rm C}^{18}{\rm O}\) intensity
and the adopted physical conversion, not by the error prescription.

The \(^{13}{\rm C}^{17}{\rm O}\) \(3\!-\!2\) detection reported by
\citet{Booth2019} provides a complementary constraint from an even rarer CO
isotopologue. Their full-disk radiative-transfer model, which included the blended
hyperfine multiplet, required an increase in the global CO gas mass by a factor of
3.5 to reproduce the observed emission,
while \(^{13}{\rm C}^{17}{\rm O}\) remained optically thin. A direct
numerical comparison of their optical depth with ours is not one-to-one
because the measurements involve different rotational transitions, excitation 
conditions, spatial scales, and spectral resolutions, and because 
the \(^{13}{\rm C}^{17}{\rm O}\) line strength 
is distributed among multiple blended hyperfine components.

Under comparable physical conditions, the lower \(^{13}{\rm C}^{17}{\rm O}\)
abundance and hyperfine partitioning imply a lower peak optical depth than that of
\(^{13}{\rm C}^{18}{\rm O}\) \(2\!-\!1\). The
moderate \(^{13}{\rm C}^{18}{\rm O}\) optical depths inferred here are therefore
not in tension with the optically thin \(^{13}{\rm C}^{17}{\rm O}\) result. 
The \citet{Booth2019} result nevertheless provides complementary global
evidence for a substantial CO reservoir in HD~163296.

\subsection{Beam-Forward Consistency Check and Transition Morphology}
\label{subsec:beamforward_results}

The local inversion provides the most transparent abundance constraint, but it
does not explicitly account for radial beam coupling. We therefore compare it
with the beam-forward radial-profile MCMC described in
Section~\ref{subsec:beamforward_consistency}. In this calculation, a flexible
PCHIP \(N_{\rm CO}(R)\) nuisance profile is converted to
\(^{13}{\rm C}^{18}{\rm O}\) intensity with the same \code{RADEX} setup, projected and
beam-convolved, and fitted to the observed radial profile.

Across the same eight fixed nuisance-parameter cases, the beam-forward MCMC
gives
\[
N_{\rm CO}^{\rm bf}(70\text{--}75~{\rm au})
=
(1.5\text{--}2.6)\times10^{20}~{\rm cm^{-2}},
\]
with
\[
\tau_{^{13}{\rm C}^{18}{\rm O}(2-1)}
=
0.36\text{--}1.06 .
\]
These values agree with the direct beam-averaged inversion within the tested
systematic range, indicating that radial beam coupling does not change the
70--75 au abundance conclusion.

The beam-forward profiles also provide a descriptive view of the broader
CO-column transition. They consistently place 70--75 au on the high-column
side and show a rapid outward decline over the following \(\sim20\)--30 au,
broadly centered near \(\sim80\)--90 au. We interpret this as the
CO-column manifestation of the snowline transition. However, because the beam
FWHM is comparable to the transition scale, and because the profiles remain
conditioned on the adopted temperature structure, linewidth, isotope ratios,
and smoothness prior, we do not treat the fitted midpoint, width, or contrast
as fully resolved, model-independent snowline parameters.

\subsection{Physical Column Budget}
\label{subsec:physical_budget}
As a physical consistency check, the adopted three-dimensional
disk structure described in Section~\ref{subsec:physical_model} is used to 
determine whether sufficient gas resides in the
warm, shielded CO-bearing phase to account for the empirically inferred
\(N_{\rm CO}\) near the snowline.
For each radius, we integrate the gas-phase CO abundance over the vertical
structure, $
N_{\rm CO}^{\rm phys}(R)=\int X_{\rm CO}(R,z)\,n_{\rm H2}(R,z)\,dz$.
We adopt \(X_{\rm CO}=7.5\times10^{-5}\) in the warm molecular layer (close to the canonical dense cloud abundance \(\sim10^{-4}\)), together with
interstellar isotope ratios, a CO freezeout temperature of 25 K, and a
freezeout reduction factor of 320 below this temperature
\citep{QiWilner2024}. The upper boundary of the CO layer is set by
an adopted
photodissociation shielding threshold,
\(N_{\rm H_2}=1.6\times10^{21}\,{\rm cm^{-2}}\), measured vertically
downward from the disk surface
\citep[e.g.,][]{vanZadelhoff2003,AikawaNomura2006,Qi2011,QiWilner2024}.
This value is a parametric approximation to the CO photodissociation
boundary.
We evaluate two versions of
the physical model: one using the smooth MAPS gas surface density and one
including Gaussian gas gaps. For comparison with the beam-averaged empirical
columns, the model columns are radially smoothed to approximate the effect of
the synthesized beam.

\begin{figure*}[ht!]
\centering
\includegraphics[width=\linewidth]{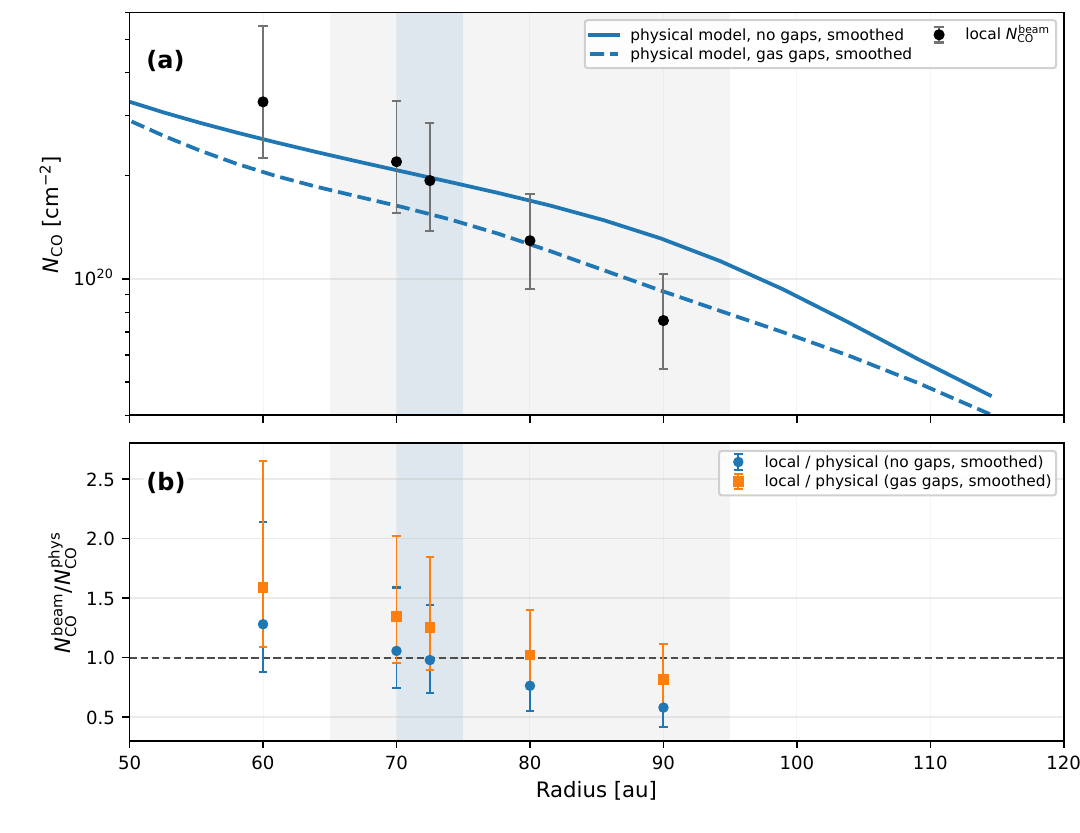}
\caption{
Physical-model CO-column budget. Top: CO columns computed from the physical
disk model with and without Gaussian gas gaps, compared with the empirical
local \(N_{\rm CO}^{\rm beam}\) measurements. The physical model
curves are radially smoothed to approximate beam averaging. Bottom: ratio of
the empirical local columns to the corresponding physical model predictions. 
The broad light gray region marks the profile steepening interval identified in 
Figures~\ref{fig:profile_derivative} and \ref{fig:beamavg_nco_tau}, while the blue region
marks the fiducial 70--75 au annulus on its inner, high column side. Both
physical models provide CO columns of order
\(10^{20}\,{\rm cm^{-2}}\) near this radius and are consistent with the
empirical constraint within its uncertainties.
}
\label{fig:physical_column_compare}
\end{figure*}

Figure~\ref{fig:physical_column_compare} compares the empirical local CO columns with
those predicted by two variants of the physical disk model. The fiducial annulus lies on the inner, high-column side of the broader profile-steepening region identified from the observed radial profile
and its derivative. The physical model was constructed independently of the 
present  \(^{13}{\rm C}^{18}{\rm O}\) inversion and was
not fitted to the observed \(^{13}{\rm C}^{18}{\rm O}\)  radial profile. It is used
here only to test whether the adopted gas and temperature structures, together
with a near-canonical warm CO abundance, can supply the inferred local column.

Averaged over 70--75 au,
the beam-smoothed columns are \(N_{\rm CO}^{\rm phys}=2.0\times10^{20}\,{\rm cm^{-2}}\)
for the smooth gas model and \(1.5\times10^{20}\,{\rm cm^{-2}}\) for the model with
gas gaps. Both are comparable to the empirical
\(N_{\rm CO}^{\rm beam}=(1.6\text{--}2.4)\times10^{20}\,{\rm cm^{-2}}\).
Thus, under the adopted gas, thermal, and chemical prescriptions, the
model places sufficient material in the warm, shielded CO-bearing layer
to account for the measured local CO column without requiring an
order-of-magnitude reduction in the assumed warm-layer abundance. This
comparison is a column-budget consistency check rather than an
independent determination of \(X_{\rm CO}\). It is not intended to
reproduce the detailed radial morphology of the snowline transition or
to provide a unique physical interpretation of it.

\section{Discussions}
\label{sec:discussions}

\subsection{Linewidth, Optical Depth, and Abundance Systematics}
\label{subsec:systematics}

The dominant radiative-transfer systematic is the local linewidth used in the
\code{RADEX} conversion between \(N_{\rm CO}\) and emergent
\(^{13}{\rm C}^{18}{\rm O}\) intensity. The
\(\Delta v=0.30~{\rm km~s^{-1}}\) cases give larger columns and higher
line-center optical depths, with
\(\tau_{^{13}{\rm C}^{18}{\rm O}}\simeq0.78\)--0.97 in the fiducial
70--75 au beam-averaged inversion, whereas the
\(\Delta v=0.50~{\rm km~s^{-1}}\) cases give
\(\tau_{^{13}{\rm C}^{18}{\rm O}}\simeq0.39\)--0.46. This behavior is
expected for a marginally optically thick line: increasing the local linewidth
reduces the line-center optical depth and changes the opacity correction
required to match the observed integrated intensity. The resulting range
shows that \(^{13}{\rm C}^{18}{\rm O}\)
\(2\!-\!1\) is marginally optically thick and therefore cannot be treated as safely optically thin at the snowline edge.

A Keplerian aligned annular stack gives an effective FWHM of
\(\sim1.2~{\rm km~s^{-1}}\) near 70--75 au, but this width should not be
identified with the local microphysical linewidth used in \code{RADEX}. A beam-shear
estimate for the observed \(0\farcs274\times0\farcs246\) beam gives a
comparable width from unresolved Keplerian velocity gradients alone. We
therefore treat \(\Delta v=0.30\)--0.50 km s\(^{-1}\) as a physically
motivated bracket for the local linewidth rather than as a measurement of turbulent broadening.

We also tested the sensitivity of the local column inference to JvM residual
scaling. As described in Appendix~\ref{app:jvm_test}, the cube without JvM
residual scaling gives a moderately higher integrated intensity and consequently,
larger inferred CO columns and line-center optical depths. Although the difference is measurable, 
it does not alter the conclusion that the 70--75 au annulus contains a high local column and that 
\(^{13}{\rm C}^{18}{\rm O}\) \(2\!-\!1\) has non-negligible opacity.

Dust opacity constitutes an additional systematic that is not included in
the local slab calculation. Continuum subtraction removes the measured
continuum level but does not restore line photons absorbed or scattered by
dust. If a significant fraction of the
\(^{13}{\rm C}^{18}{\rm O}\) emission originates behind or within an
optically non-negligible dust layer, the observed continuum-subtracted
intensity will underestimate the intrinsic line emission. Neglecting this
effect would therefore tend to bias \(N_{\rm CO}^{\rm beam}\) low,
although the magnitude depends on the relative vertical distributions and
temperatures, and optical depths of the gas and dust. A coupled line and dust radiative transfer
calculation would be required to quantify this correction.

The preceding forward calculation tests whether an adopted three-dimensional
disk structure and fixed warm-layer CO abundance can supply the observed
gas-phase CO column. A separate inverse step is required to convert the
empirically inferred \(N_{\rm CO}^{\rm beam}\) into a CO/H\(_2\) ratio:
the measured CO column is divided by an adopted total H\(_2\) column. We
denote this column-averaged gas-phase ratio as
\(
\overline{X}_{\rm CO}^{\rm col}
\equiv
\frac{N_{\rm CO}^{\rm beam}}{N_{\rm H2}} .
\)
This quantity is distinct from the local warm-layer abundance
\(X_{\rm CO}\) adopted in the physical model and introduces a systematic
associated with the \(N_{\rm H2}\) denominator.

As a reference, we adopt the HD~163296 gas surface-density profile used by
\citet{Zhang2021} and \citet{Armitage2026},
\(
\Sigma_{\rm gas}(R)
=
\Sigma_c
\left(\frac{R}{R_c}\right)^{-\gamma}
\exp\left[
-\left(\frac{R}{R_c}\right)^{2-\gamma}
\right],
\)
with \(R_c=165~{\rm au}\), \(\gamma=0.8\), and
\(\Sigma_c=8.8~{\rm g~cm^{-2}}\). This profile gives
\(\Sigma_{\rm gas}\simeq11.2\)--\(12.2~{\rm g~cm^{-2}}\) across
70--75 au. Using
\(\Sigma_{\rm gas}=2.8m_{\rm H}N_{\rm H2}\), where the factor 2.8
accounts for helium and heavy elements per H\(_2\) molecule
\citep[e.g.,][]{Kauffmann2008,Draine2011}, this corresponds to
\(N_{\rm H2}\simeq(2.4\)--\(2.6)\times10^{24}~{\rm cm^{-2}}\).
We adopt \(N_{\rm H2}=2.5\times10^{24}~{\rm cm^{-2}}\) as the
reference value. This quantity is not a direct local measurement of
\(N_{\rm H2}\), but follows from the parametric gas surface-density profile
adopted by \citet{Zhang2021} in a thermochemical model constrained jointly
by the SED, millimeter continuum, and CO emission surfaces. Because the same
gas surface-density normalization enters the smooth physical model in
Section~\ref{subsec:physical_budget}, the resulting abundance conversion is not
an independent validation of the forward column-budget calculation.

An external estimate of the gas column normalization is available from the rotation curve
analysis of \citet{Martire2024}. Their vertically stratified fit to the
\(^{12}{\rm CO}\) and \(^{13}{\rm CO}\) rotation curves gives
\(M_{\rm disk}=0.134\pm0.001\,M_\odot\) and
\(R_c=91\pm1\) au. Evaluating their adopted self-similar surface density
profile gives
\(\Sigma_{\rm gas}\simeq12.2\)--\(13.8~{\rm g~cm^{-2}}\) over
70--75 au, corresponding to
\(N_{\rm H2}\simeq(2.6\)--\(2.9)\times10^{24}~{\rm cm^{-2}}\).
This independently constructed parametric model gives a hydrogen column
close to the MAPS-based reference value and therefore supports the
adopted denominator. It is nevertheless not a nonparametric measurement of
the individual annulus and remains sensitive to the assumed thermal
structure and emitting surfaces.

Our auxiliary continuum modeling provides a further plausibility check. It gives an annulus-averaged dust surface density of $\Sigma_{\rm dust}\simeq4\times10^{-2}~{\rm g~cm^{-2}}$ over the 58--85 au region, a factor of $\sim2$--3 lower than the standard grain model of \citet{Guidi2022}. Such differences are within the systematic uncertainties of continuum-based dust-mass estimates, which depend strongly on the adopted dust opacity model \citep[e.g.,][]{Birnstiel2018, Zagaria2025,Ueda2025}. 
%For the reference gas column, $N_{\rm H2}=2.5\times10^{24}~{\rm cm^{-2}}$, 
Combined with the reference gas column, this dust surface density implies a local gas-to-dust mass ratio of $\sim300$, within a factor of a few of the interstellar value. Because the dust surface density is model-dependent and
averaged over a wider radial interval than the fiducial CO annulus, this
comparison is intended only as an approximate plausibility check.

For the reference \(N_{\rm H2}=2.5\times10^{24}~{\rm cm^{-2}}\), the
fiducial 70--75 au beam-averaged posterior medians imply
\(\overline{X}_{\rm CO}^{\rm col}={\rm CO/H_2}
=(6.4\text{--}9.5)\times10^{-5}\). More generally,
\({\rm CO/H_2}=8.0\times10^{-5}
\left(N_{\rm CO}^{\rm beam}/2.0\times10^{20}~{\rm cm^{-2}}\right)
\left(N_{\rm H2}/2.5\times10^{24}~{\rm cm^{-2}}\right)^{-1}\).
Using the representative kinematic column from \citet{Martire2024} 
instead gives approximately
\({\rm CO/H_2}=(5.7\text{--}8.6)\times10^{-5}\), leaving the
interpretation unchanged. A wider systematic range could extend to
\(N_{\rm H2}\sim2\times10^{24}\)--\(10^{25}~{\rm cm^{-2}}\), although
the upper end reflects dust model degeneracies rather than either of the
preferred gas surface density estimates. The most direct quantitative results
of this study are therefore \(N_{\rm CO}^{\rm beam}\) and
\(\tau_{^{13}{\rm C}^{18}{\rm O}}\). The column averaged gas phase CO/H\(_2\) value remains
conditional on the adopted local H\(_2\) column.

\subsection{Comparison with Previous Constraints and Implications for CO Depletion}
\label{subsec:co_depletion_implications}
 
The most relevant previous resolved comparison is the C\(^{17}\)O
\(1\!-\!0\) analysis of \citet{Zhang2021}, obtained at a spatial resolution comparable to that of the present \(^{13}{\rm C}^{18}{\rm O}\) observations. Their C\(^{18}\)O
\(2\!-\!1\) based profiles give
\(N_{\rm CO}\simeq1.7\)--\(2.2\times10^{19}~{\rm cm^{-2}}\) over
70--75 au, approximately an order of magnitude below the beam averaged
column inferred here. However, both their thermochemical C\(^{17}\)O
retrieval and their direct hyperfine fit give substantially higher columns
between approximately 30 and 150 au. Inside approximately 100 au, the
C\(^{17}\)O based columns are higher than the C\(^{18}\)O based values
by factors of approximately 2--6. The C\(^{17}\)O result also gives the
better match to the then available disk integrated
\(^{13}{\rm C}^{18}{\rm O}\) spectrum, and \citet{Zhang2021} adopted
the C\(^{17}\)O based column inside 150 au in their preferred gas mass
calculation.

Our present \(N_{\rm CO}^{\rm beam}=(1.6\)--\(2.4)
\times10^{20}~{\rm cm^{-2}}\) measurement is consistent with the higher
MAPS C\(^{17}\)O branch. It provides an independent and more spatially
localized test using the absolute brightness of a rarer isotopologue at the
resolved snowline edge. The C\(^{17}\)O hyperfine analysis remains an
important empirical constraint, although \citet{Zhang2021} noted that the fits become unreliable inside \(\sim45\) au because of complete blending of the hyperfine components and beam smearing. The finite spectral resolution also blends two of the three hyperfine components throughout the disk. At our 70--75 au fiducial annulus, these innermost limitations are less severe, although C\(^{17}\)O remains more optically thick than
\(^{13}{\rm C}^{18}{\rm O}\).

The \(^{13}{\rm C}^{17}{\rm O}\) \(3\!-\!2\) detection by
\citet{Booth2019} provides another important indication of a substantial CO
reservoir in HD~163296. Their radiative transfer model underpredicted the
disk integrated \(^{13}{\rm C}^{17}{\rm O}\) intensity and required a
factor of 3.5 increase in the global CO gas mass. They also showed that
C\(^{18}\)O is optically thick inside the snowline. That result is consistent
with the high column inferred here, but it does not provide a direct local
measurement at 70--75 au because the \(^{13}{\rm C}^{17}{\rm O}\)
constraint was spatially and spectrally unresolved and was interpreted through
a global disk model. Rather than representing the first evidence for a high CO column in this disk,
The present measurement should be viewed as a
resolved local confirmation and refinement of the substantial CO reservoir indicated by \citet{Booth2019} and by the MAPS C\(^{17}\)O analysis.

The corresponding abundance is also consistent with earlier model based
results. \citet{Zhang2021} found that CO/H\(_2\) rises inward across the
HD~163296 snowline and approaches the interstellar value when constrained
with C\(^{17}\)O. More recently, \citet{Trapman2025} fitted integrated
continuum, \(^{13}{\rm CO}\), C\(^{18}\)O, and N\(_2\)H\(^+\) fluxes
with a thermochemical model grid. Their reported
\(\log_{10}x_{\rm CO}=-4.28^{+0.17}_{-0.21}\) for HD~163296 is a
global model parameter describing the peak CO abundance in the emitting
layer, rather than a disk averaged abundance or a resolved measurement at a
specific radius. Its value, approximately \(5.2\times10^{-5}\), is
nevertheless consistent with the conditional local abundance inferred here.
Their broader result that Herbig disks have near interstellar peak CO
abundances, while T~Tauri disks are commonly inferred to have lower values,
also supports the conclusion that strong warm gas CO depletion is not
universal.

These agreements do not make the approaches equivalent. The MAPS,
\citet{Booth2019}, and \citet{Trapman2025} results depend on global disk
or thermochemical models, whereas the present analysis first extracts the
local empirical information available from the resolved
\(^{13}{\rm C}^{18}{\rm O}\) emission. In particular, we constrain the
absolute beam averaged CO column and show explicitly that
\(^{13}{\rm C}^{18}{\rm O}\) \(2\!-\!1\) has
\(\tau\simeq0.39\)--0.97 at 70--75 au. Even this rare tracer is therefore
not safely optically thin at the snowline edge and is expected to become more
opaque farther inward. This limits any direct conversion of the inner radial
intensity profile into an abundance profile without an explicit treatment of
opacity, excitation, and beam coupling.

There is also a distinction between establishing an inward rise in
\(N_{\rm CO}\) and assigning that rise to a particular physical mechanism.
\citet{Zhang2021} interpreted the higher inner abundance in the context of
settling, inward pebble drift, and sublimation of CO rich ices.
\citet{Armitage2026}, using the same resolved
\(^{13}{\rm C}^{18}{\rm O}\) data analyzed here, fitted a full
thermochemical and radiative transfer model with a radially varying CO
abundance and inferred a strongly centrally concentrated enhancement
associated with pebble drift. By contrast, \citet{QiWilner2024} showed that
the vertical temperature structure can itself generate a sharp change in the
warm, shielded CO column across the snowline.

The thermal snowline transition and pebble driven enrichment are not mutually
exclusive. The relevant distinction is the order of inference. The observed
brightness, opacity, and local CO column should first be established before the
inward rise is assigned to transport or chemical evolution through a detailed
physical model. A sharp increase in \(N_{\rm CO}\) toward smaller radii can
result from the rapid expansion of the warm, shielded CO layer across the
snowline, while pebble drift may contribute an additional abundance
enhancement farther inward. The present measurement supplies a local empirical
anchor for separating these effects, but it does not by itself confirm or rule out pebble drift in the inner disk.

The inferred abundance should therefore be interpreted as a local constraint
on the high column side of the snowline transition, not as evidence that CO is
undepleted throughout HD~163296. The outer disk, the low column side of the
transition, and the disk atmosphere may still have reduced gas phase CO because
of freezeout, photochemistry, grain surface conversion, or sequestration into
larger solids. At 70--75 au, however, an abundance as low as
\({\rm CO/H_2}\sim10^{-5}\) would require
\(N_{\rm H2}\gtrsim(1.6\)--\(2.4)\times10^{25}~{\rm cm^{-2}}\), well
above both the adopted MAPS column and the independent kinematic estimate from
\citet{Martire2024}. The absolute \(^{13}{\rm C}^{18}{\rm O}\)
intensity therefore does not require order of magnitude CO depletion on the
warm side of the resolved snowline transition.

%\subsection{Toward a Fully Empirical CO Column Profile}
\subsection{Limitations and Prospects for an Empirical CO Column Profile}
\label{subsec:future_empirical_nco}

The present analysis provides a local beam-averaged constraint on
\(N_{\rm CO}\) and
\(\tau_{^{13}{\rm C}^{18}{\rm O}}\) on the high-column side of the
snowline transition, together with a beam-forward characterization of the
broader radial decline. It does not yield a unique
\(N_{\rm CO}(R)\) reconstruction. Such a reconstruction from a single
CO-isotopologue transition remains sensitive to the excitation temperature,
optical depth, local linewidth, emitting geometry, and isotope ratios.
These quantities are constrained or bracketed here, but are not all
determined by the \(^{13}{\rm C}^{18}{\rm O}\) \(2\!-\!1\) data alone.

A more empirical \(N_{\rm CO}(R)\) reconstruction will require
matched-resolution observations of multiple isotopologues and transitions
across the CO snowline. Same-transition ratios of rare isotopologues, such as
\(^{13}{\rm C}^{18}{\rm O}\)/C\(^{17}\)O \(2\!-\!1\) or
C\(^{18}\)O/C\(^{17}\)O \(1\!-\!0\), can constrain optical depth and
isotopic conversion factors while minimizing excitation differences.
Multiple transitions of the same rare isotopologue can constrain
\(T_{\rm ex}(R)\) and test whether the emission is thermalized. The
observations must also resolve the transition. The present beam FWHM is
\(\simeq26\) au, so neighboring annuli in the beam-forward reconstruction
are strongly correlated. Imaging the rare isotopologues at
\(\lesssim10\)--15 au resolution with sufficient sensitivity would be
needed to constrain the intrinsic transition shape. Such angular resolution is readily achievable with ALMA; the principal
observational challenge is obtaining sufficient surface-brightness
sensitivity in these intrinsically weak lines.

Even with such data, converting \(N_{\rm CO}(R)\) into CO/H\(_2\) would
require an independent hydrogen-column constraint. Dust-based gas columns
depend on the opacity, scattering, grain-size distribution, and assumed
dust-to-gas ratio, whereas CO-based gas columns are degenerate with the CO
abundance being inferred. The most direct next step is therefore a
matched-resolution, multi-isotopologue reconstruction of
\(N_{\rm CO}(R)\); any corresponding CO/H\(_2\) profile will remain
conditional on the adopted gas-column constraint.

\section{Summary and Conclusions}
\label{sec:summary}

We have used spatially resolved
\(^{13}{\rm C}^{18}{\rm O}\) \(2\!-\!1\) emission toward HD~163296 to
constrain the CO column and line optical depth on the high-column side of
the resolved CO snowline transition. The principal results are summarized
as follows.

\begin{enumerate}

\item The radial integrated-intensity profile decreases sharply near the
CO snowline, with a local minimum in its radial derivative at
\(\simeq0\farcs75\)--\(0\farcs79\). We adopt the 70--75 au annulus,
located on the inner, high-column side of this transition, for the local
column analysis.

\item Converting the absolute beam-averaged
\(^{13}{\rm C}^{18}{\rm O}\) \(2\!-\!1\) intensity with
\code{RADEX} gives
\[
N_{\rm CO}^{\rm beam}
=
(1.6\text{--}2.4)\times10^{20}~{\rm cm^{-2}}
\]
over 70--75 au. The range brackets the adopted local linewidths and
isotope-ratio pairs. The inferred column is insensitive to the gas density
over the range relevant to the emitting region.

\item The corresponding line-center optical depth is
\[
\tau_{^{13}{\rm C}^{18}{\rm O}}
=
0.39\text{--}0.97 .
\]
The rare \(^{13}{\rm C}^{18}{\rm O}\) \(2\!-\!1\) line is therefore
marginally optically thin to moderately opaque at the snowline edge and
cannot be treated safely in the optically thin limit. The dominant
radiative-transfer systematic is the local linewidth. Tests of JvM
residual scaling change the inferred intensity, column, and opacity
moderately but do not alter this conclusion.

\item A beam-forward reconstruction confirms that a pronounced radial
decline in the intrinsic CO column is required to reproduce the observed
profile. Because neighboring radial elements are strongly coupled by the
\(\simeq26\) au beam and by the adopted regularization, this reconstruction
is interpreted as a consistency check on the broader transition rather
than as a unique deconvolution of \(N_{\rm CO}(R)\).

\item A forward physical column-budget calculation, based on the adopted
gas density and temperature structures together with a near-canonical
warm-layer CO abundance, gives beam-smoothed columns of
\(2.0\times10^{20}~{\rm cm^{-2}}\) for the smooth gas model and
\(1.5\times10^{20}~{\rm cm^{-2}}\) for the model with gas gaps over
70--75 au. The adopted physical structure can therefore supply the
empirically inferred local column without requiring an order-of-magnitude
reduction in the warm CO abundance. This calculation is not a fit to the
observed radial profile and does not provide a unique physical explanation
for the snowline morphology.

\item Adopting a reference hydrogen column of
\(N_{\rm H2}=2.5\times10^{24}~{\rm cm^{-2}}\) gives a column-averaged
gas-phase abundance
\[
{\rm CO/H_2}
=
(6.4\text{--}9.5)\times10^{-5}.
\]
A separate surface-density estimate from the stratified rotation-curve
analysis gives a similar hydrogen column and hence a comparable abundance.
Nevertheless, \(N_{\rm H2}\) is obtained from parametric disk models rather
than measured directly in the individual annulus. The CO/H\(_2\) value is
therefore conditional on the adopted hydrogen-column normalization.

\end{enumerate}

The most direct quantitative results of this analysis are the local
beam-averaged \(N_{\rm CO}\) and
\(\tau_{^{13}{\rm C}^{18}{\rm O}}\). These results show that the inner side
of the resolved snowline transition contains a substantial gas-phase CO
column and that even this rare isotopologue has non-negligible opacity.
The sharp transition and high local column can arise within a plausible
thermal and density structure and do not, by themselves, uniquely require
either a warm-side abundance enhancement or inward pebble delivery,
although such transport may still contribute to the inner-disk CO
distribution. Matched-resolution observations of multiple rare
isotopologues and transitions will be required to reconstruct a more
empirical \(N_{\rm CO}(R)\) profile and to separate excitation, opacity,
and abundance effects across the snowline.

%% Please use the acknowledgment and contribution environments. This will 
%% be anonomyized when the "anonymous" style option is used. 
\begin{acknowledgments}
We thanks Anna Fehr and Sean Andrews for providing the HD~163296 temperature structure table, which were used to condition the local \code{RADEX} excitation calculations. 
T.U. acknowledges support from JSPS Overseas Research Fellowships. 
This paper makes use of the following ALMA data: ADS/JAO.ALMA\#2021.1.00899.S. ALMA is a partnership of ESO (representing its member states), NSF (USA), and NINS (Japan), together with NRC (Canada), NSTC and ASIAA (Taiwan), and KASI (Republic of Korea), in cooperation with the Republic of Chile. The Joint ALMA Observatory is operated by ESO, AUI/NRAO and NAOJ.
The National Radio Astronomy Observatory is a facility of the National 
Science Foundation operated under a cooperative agreement by Associated Universities, Inc.
\end{acknowledgments}

%% To help institutions obtain information on the effectiveness of their 
%% telescopes the AAS Journals has created a group of keywords for telescope 
%% facilities.
%
%% Following the acknowledgments section, use the following syntax and the
%% \facility{} or \facilities{} macros to list the keywords of facilities used 
%% in the research for the paper.  Each keyword is check against the master 
%% list during copy editing.  Individual instruments can be provided in 
%% parentheses, after the keyword, but they are not verified.
\facilities{ALMA}

%% Similar to \facility{}, there is the optional \software command to allow 
%% authors a place to specify which programs were used during the creation of 
%% the manuscript. Authors should list each code and include either a
%% citation or url to the code inside ()s when available.
\software{Astropy \citep{Astropy2013,Astropy2018,Astropy2022}, \code{CASA} \citep{CASA+2022},  \code{emcee} \citep{emcee}, \code{GoFish} \citep{Teague2019}, Matplotlib \citep{Hunter2007}, NumPy \citep{vanderWalt2011}, \code{RADEX} \citep{vanderTak2007}, \code{SpectralRadex} \citep{Holdship2021}, \code{RADMC-3D} \citep{RADMC}.}         

%% Appendix material should be preceded with a single \appendix command.
%% There should be a \section command for each appendix. Mark appendix
%% subsections with the same markup you use in the main body of the paper.
%%
%% Each Appendix (indicated with \section) will be lettered A, B, C, etc.
%% The equation counter will reset when it encounters the \appendix
%% command and will number appendix equations (A1), (A2), etc. The
%% Figure and Table counter will not reset.

\appendix
\section{Beam-Forward Radial-Profile Consistency Check}
\label{app:beamforward_profiles}

\begin{figure}[ht!]
\centering
\includegraphics[width=0.95\textwidth]{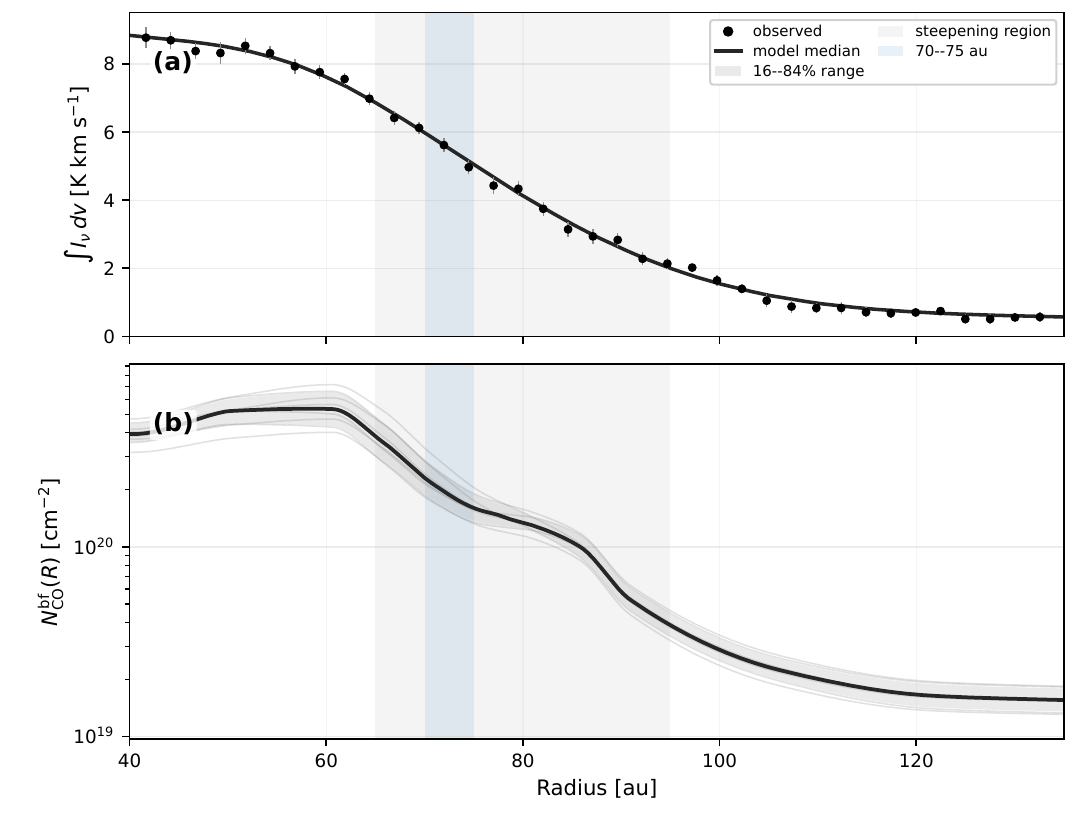}
\caption{
Beam-forward radial-profile consistency check.
(a) Observed \(^{13}{\rm C}^{18}{\rm O}\) \(2\!-\!1\) radial integrated
intensity profile compared with the beam-convolved model profiles from the
fixed nuisance-parameter cases. The agreement shows that the flexible
beam-forward profiles reproduce the observed absolute intensity profile used
in the local inversion.
(b) Corresponding model-conditioned intrinsic CO-column profiles,
\(N_{\rm CO}^{\rm bf}(R)\). Thin curves show the fixed nuisance-parameter
cases, and the thick curve or shaded region shows their median/systematic
range. The blue band marks the fiducial 70--75 au annulus; the gray band marks
the profile-steepening region identified from the observed derivative. The
profiles place the fiducial annulus on the high-column side of a rapid outward
CO-column decline, but the detailed intrinsic profile is not interpreted as a
unique model-independent \(N_{\rm CO}(R)\) measurement.
}
\label{fig:beamforward_profiles}
\end{figure}

Figure~\ref{fig:beamforward_profiles} shows the beam-forward radial-profile
fits used as a consistency check on the local beam-averaged inversion. The
model profiles reproduce the observed beam-convolved
\(^{13}{\rm C}^{18}{\rm O}\) \(2\!-\!1\) radial intensity while allowing the
surrounding CO-column structure to vary. The corresponding intrinsic
\(N_{\rm CO}^{\rm bf}(R)\) profiles consistently place the 70--75 au annulus
on the high-column side of a rapid outward decline. This supports the
interpretation that the local beam-averaged column in the main text samples
the high-column side of the CO snowline transition.

These profiles should not be interpreted as a unique, model-independent
empirical reconstruction of \(N_{\rm CO}(R)\). The detailed intrinsic shape is
conditioned on the adopted Fehr--Andrews temperature structure, local
linewidth, isotope ratios, emitting height, and weak smoothness prior, and the
transition scale is comparable to the \(\simeq26\) au radial beam. We therefore
use the beam-forward profiles only to test radial beam-coupling effects and to
provide a descriptive view of the broader CO-column transition.

\section{Sensitivity to JvM Residual Scaling}
\label{app:jvm_test}
 
The JvM correction rescales the residual image to account for the different
areas of the dirty and restoring beams \citep{Jorsater1995,
Casassus2022}. Because this correction can affect faint emission that
remains in the residual image, we repeated the analysis using the standard
restored, primary-beam-corrected cube without JvM residual scaling. The JvM
and non-JvM products were constructed from the same visibilities, weighting,
channelization, CLEAN model, restoring beam, CLEAN mask, and stopping
threshold. We then repeated the GoFish profile extraction using the same disk
geometry, Keplerian mask, velocity interval, and radial bins.

\begin{figure}[ht!]
\centering
\includegraphics[width=\linewidth]
{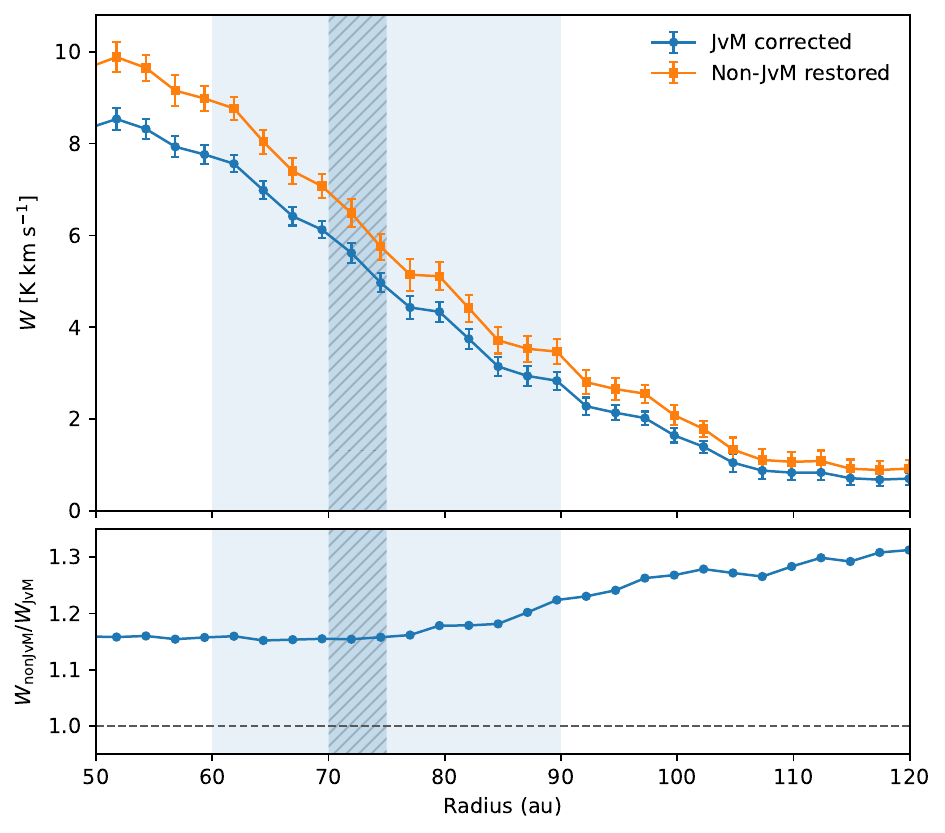}
\caption{
Sensitivity of the \(^{13}{\rm C}^{18}{\rm O}\) \(2\!-\!1\) radial
profile to JvM residual scaling. Top: integrated-intensity profiles extracted
from the JvM-corrected and standard restored, non-JvM cubes using otherwise
identical analysis settings. Bottom: ratio of the non-JvM to JvM-corrected
profiles. The horizontal dashed line marks equality. The broad light gray
region marks the profile-steepening interval identified in
Figures~\ref{fig:profile_derivative}, and the blue hatched
region marks the 70--75 au fiducial annulus. The non-JvM profile is
approximately \(15\%\) brighter in the fiducial annulus, while the radial
morphology is preserved.
}
\label{fig:jvm_comparison}
\end{figure}

Figure~\ref{fig:jvm_comparison} compares the resulting radial profiles. The
non-JvM profile preserves the location and overall morphology of the profile
steepening but is approximately \(15\%\) brighter over the 70--75 au
fiducial annulus. The difference increases gradually at larger radii, where a
larger fraction of the emission is likely retained in the residual image.

We propagated the non-JvM profile through the same local \code{RADEX}
inversions used for the fiducial analysis, including linewidths of 0.30 and
0.50 km s\(^{-1}\) and both the ISM and QiMed isotope-ratio sets.
The results are summarized in Table~\ref{tab:jvm_comparison}. The non-JvM
cube gives
\(N_{\rm CO}^{\rm beam}=(2.0\)--\(3.2)\times10^{20}\,{\rm cm^{-2}}\)
and
\(\tau_{^{13}{\rm C}^{18}{\rm O}}=0.57\)--1.29, compared with
\((1.6\)--\(2.4)\times10^{20}\,{\rm cm^{-2}}\) and
\(\tau=0.39\)--0.97 for the JvM-corrected cube. Residual scaling therefore
introduces a measurable quantitative systematic, but both image products
require a high local CO column and show that
\(^{13}{\rm C}^{18}{\rm O}\) \(2\!-\!1\) is not safely optically thin at
the snowline edge. We retain the JvM-corrected result as the fiducial
measurement and treat the non-JvM result as an imaging systematic.

\begin{table}[ht!]
\centering
\caption{Sensitivity of the 70--75 au inference to JvM residual scaling.}
\label{tab:jvm_comparison}
\begin{tabular}{lccc}
\hline
\hline
Image product &
\(\langle W\rangle/\langle W\rangle_{\rm JvM}\) &
\(N_{\rm CO}^{\rm beam}\) &
\(\tau_{^{13}{\rm C}^{18}{\rm O}}\) \\
&
&
(\(10^{20}\,{\rm cm^{-2}}\)) &
\\
\hline
JvM corrected     & 1.00          & 1.6--2.4 & 0.39--0.97 \\
Standard restored & \(\simeq1.15\) & 2.0--3.2 & 0.57--1.29 \\
\hline
\end{tabular}

\tablecomments{
The quoted ranges span linewidths of 0.30 and 0.50 km s\(^{-1}\) and the
ISM and QiMed isotope-ratio sets. The JvM-corrected cube is adopted for the
fiducial analysis.
}
\end{table}

\bibliography{main}{}
\bibliographystyle{aasjournalv7}

%% This command is needed to show the entire author+affiliation list when
%% the collaboration and author truncation commands are used.  It has to
%% go at the end of the manuscript.
%\allauthors

%% Include this line if you are using the \added, \replaced, \deleted
%% commands to see a summary list of all changes at the end of the article.
%\listofchanges

\end{document}